\documentclass[]{spie}  
\usepackage[]{graphicx}
\usepackage{float}
\usepackage{amssymb,amsmath}
\usepackage{wrapfig}
\usepackage{sidecap}

\title{InnoPOL: an EMCCD imaging polarimeter and 85-element curvature AO system on the 3.6-m AEOS telescope for cost effective polarimetric speckle suppression}

\author{David Harrington\supit{a,b,d}, Svetlana Berdyugina\supit{b}, Mark Chun\supit{c}, Christ Ftaclas\supit{a}, Daniel Gisler\supit{b}, Jeff Kuhn\supit{d}\skiplinehalf
\small{
\supit{a}Institute for Astronomy, University of Hawaii, 2680 Woodlawn Drive, Honolulu, HI, USA, 96822; \\
\supit{b}Kiepenheuer Institute for Solar Physics, Schoneck str. 6, Freiburg Im Breisgau, Germany D-79104; \\
\supit{c}Institute for Astronomy, University of Hawaii, 640 North A'lhoku Place, Hilo, HI, USA, 96720; \\
\supit{d}Institute for Astronomy, University of Hawaii, 34 O'hia Ku Street, Makawao, HI, USA, 96768.}
}

\authorinfo{email: {\tt dharring@hawaii.edu}}

\begin{document} 
  \maketitle 


\begin{abstract}

The Hokupa'a-85 curvature adaptive optics system components have been adapted to create a new AO-corrected coud\'{e} instrument at the 3.67m Advanced Electro-Optical System (AEOS) telescope. This new AO-corrected optical path is designed to deliver an f/40 diffraction-limited focus at wavelengths longer than 800nm. A new EMCCD-based dual-beam imaging polarimeter called InnoPOL has been designed and is presently being installed behind this corrected f/40 beam. The InnoPOL system is a flexible platform for optimizing polarimetric performance using commercial solutions and for testing modulation strategies. The system is designed as a technology test and demonstration platform as the coud\'{e} path is built using off-the-shelf components wherever possible.  Models of the polarimetric performance after AO correction show that polarization modulation at rates as slow as 200Hz can cause speckle correlations in brightness and focal plane location sufficient enough to change the speckle suppression behavior of the modulators. These models are also verified by initial EMCCD scoring camera data at AEOS. Substantial instrument trades and development efforts are explored between instrument performance parameters and various polarimetric noise sources.

\end{abstract}




\section{INTRODUCTION}

Adaptive optics combined with novel polarimetric imaging technologies is a powerful combination for detecting faint objects in the presence of overwhelming backgrounds close to bright sources. The adaptive optics correction provides a high spatial resolution image in the presence of residual atmospheric and instrument-induced speckles. By suppressing speckles via differential techniques (polarization, color, psf subtraction), targets of interest can be imaged and characterized.  The application for such techniques includes use cases for exoplanets and circumstellar material around bright stars, closely spaced objects and space situational awareness (SSA).  Speckle evolution both in brightness and focal plane position cause major limiting systematic errors. Commercial EMCCDs, liquid crystals and electronics provide fast and flexible solutions to polarimetric modulation and imaging.  Optimizing an instrument speed and performance depends critically on the scene dynamic range, field of view, atmospheric properties, and detector settings.  Several other variables complicate the design choices. 

Several astronomical instruments have combined adaptive optics systems and polarimeters with varying complexity and sophistication with a wide range of cost and design philosophies.  For instance, the SPHERE instrument on the European Southern Observatory Very Large Telescope combines an extreme adaptive optics system with a charge-shuffling broad band imaging polarimeter (ZIMPOL) for exoplanet detection with very challenging polarized point source detection criteria \cite{Beuzit:2008gt, Dohlen:2006iu, Gisler:2004ck, Roelfsema:2010ca, Thalmann:2008gi}.  The Gemini Planet Imager has a sophisticated adaptive optics system and chronograph with a single rotating retarder as a polarization modulator in a dual-beam configuration \cite{Perrin:2010dw, Macintosh:2008dn, Wiktorowicz:2012ga, Macintosh:2012eo}.  Several other projects with much lower costs deliver AO-assisted polarimetry for other astronomical use cases.  For instance, the Extreme Polarimeter (ExPo) on the 4.2m William Herschel telescope (WHT) has been developed to fit in a small instrument package using commercial detectors for fast development and deployment \cite{Rodenhuis:2008ji, Canovas:2011ed, Rodenhuis:2012du, Canovas:2012ib,Min:2012ha,Jeffers:2012ki}.

\begin{wrapfigure}{r}{0.65\textwidth}
\centering
\includegraphics[width=0.65\textwidth]{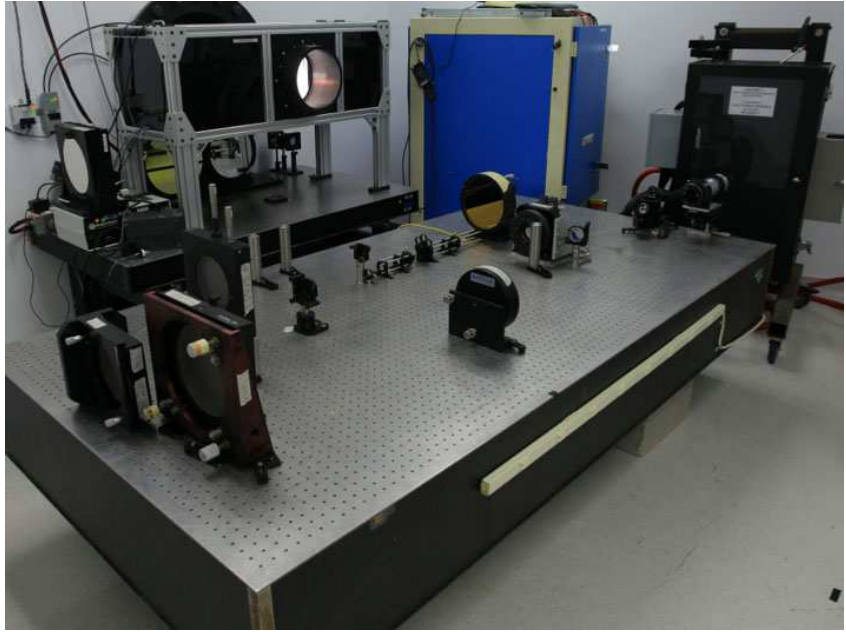}
\caption{The H-85 curvature AO system during basic installation and testing with the internal laser simulator beam. The f/200 beam enters the coud\'{e} room }
\label{h85_optical_bench_install}
 \end{wrapfigure}

\begin{figure}
\centering
\includegraphics[width=0.99\textwidth]{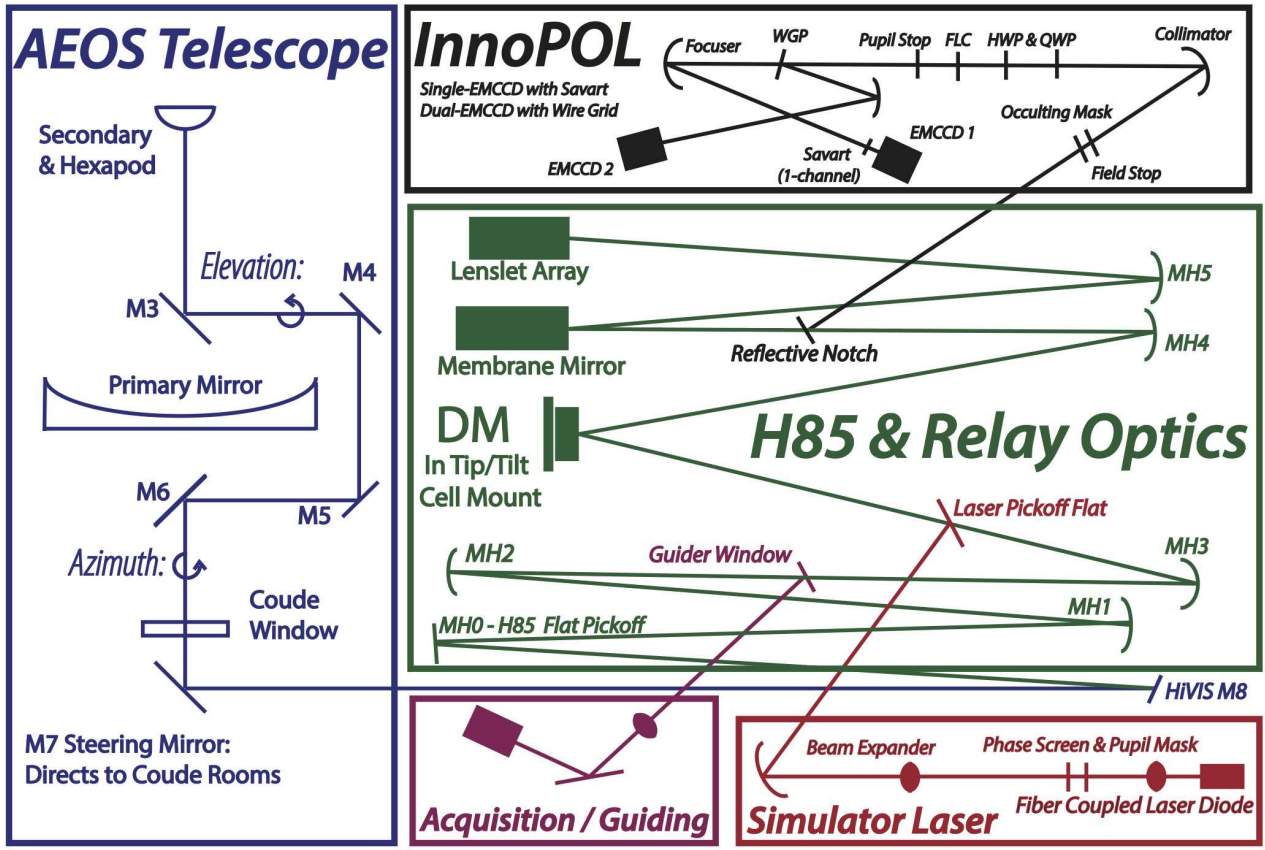}
\caption{The system optical block diagram.  Every optical element is shown and color coded according to functional group. The AEOS telescope optics in blue feed the coud\'{e} room at f/200.  The relay optics provide pathways for guiding and laser simulator injection (light \& dark red). The adaptive optics system and associated relay optics are shown in green. The H85 AO compensated f/40 beam is reflected to the InnoPOL science channel shown in black.}
\label{optical_path_layout}
\end{figure}

The 3.67m Advanced Electro-Optical System (AEOS) telescope is owned and operated by the U.S. Air Force on Haleakala, Maui, Hawai'i.  The University of Hawai'i has operated several instruments in the coud\'{e} experiment room number 3. Our new instrument developments take advantage of this room, optical relays and associated laser alignment sources.  These experiment rooms offer several advantages for technological developments.  Each coud\'{e} room is fed an f/200 beam from the telescope in a controlled laboratory environment.  Inexpensive off-the-shelf components can be used for modifications, mounting, cooling, power and other technological developments decreasing cost and allowing fast schedules. Polarimetric technologies are developing rapidly in several fields offering several novel components that could be tested with our platform in circumstellar or SSA applications \cite{Snik:2014gl, 2010amos.confE..39S}.  We combine a rebuilt Hokupa'a 85-element curvature adaptive optics system (H85) and an EMCCD-based imaging polarimeter (InnoPOL) to accomplish a few select science goals and technological developments. An optical layout for the AEOS telescope, H85 AO system, EMCCD polarimeter and associated systems is shown in Figure \ref{optical_path_layout}.  Each component will be discussed in later sections.

Atmospheric speckles nominally smooth as a statistically independent process following $\frac{1}{\sqrt{n}}$. Speckles in the AO-corrected halo are coherently modulated by the AO system itself increasing the speckles control loops lag the real atmosphere amplifying the problem with non-common path errors creating additional quasi-static speckles. \cite{1999PASP..111..587R, Hinkley:2007hf, Hinkley:2009cv, Hinkley:2011kh, Macintosh:2005ih, 2004ApJ...610L..69B, 2004ApJ...612L..85A}.  In dual beam systems, the simultaneously recorded images can be differenced subtracting residual speckles to a high degree. A dual-beam imaging polarimeter at AEOS demonstrated the speckle suppression capabilities at coud\'{e} in H-band with slow polarimetric modulation \cite{Hinkley:2009cv}. When using modulation schemes that allow sequential differencing of these normalized subtracted image pairs, most detector based noise sources are also removed (gain variations / flat fielding and several detector cosmetic issues). Non-common path errors, distortion and other optical limitations tend to dominate these residual double differenced normalized image sequences. Provided the optical path can be polarimetricaly calibrated, these large and dominant error sources can be highly suppressed. With the advent of Electron multiplying detectors with high capacity gain register pixels and frame transfer buffers, high speed polarimetric modulation is also possible to enable further options for suppressing speckle noise and optimizing dual channel imaging systems. These techniques apply to multi-color systems (such as narrow line filter differencing) as well as for polarimetric systems.  High speed polarimetric modulation also potentially allows for removing the dual-beam requirement. Such a system could use the two orthogonal polarization as independent channels for different colors, polarimetric error suppression techniques or simple increases in system efficiency.

\section{H-85 CURVATURE AO}

The Hokupa'a 85 element curvature AO system was originally designed and built as a prototype for the Gemini Near Infra-red Coronagraphic Imager (NICI) planet finding campaign \cite{2000SPIE.4007..126N, Toomey:2003cb, Chun:2008et, Liu:2010hn}. Curvature adaptive optics has provided some advantages to pursuing faint objects in SSA such as the noise-less wavefront sensing, system flexibility to adapt in changing conditions as well as a convenient platform for upgrades to both the hardware and control systems \cite{Ftaclas:2008ut, 2010amos.confE..52F, Kellerer:2010ck, Kellerer:2010gw, Mateen:2010fj}. 

The core H85 curvature AO components including the wavefront sensor (lenslet array, membrane mirror, avalanche photo diode array), deformable mirror (DM) and all associated control systems were reused in a new instrument design for the AEOS f/200 beam in coud\'{e} experiment room 3. The optical design was required to use only off-the shelf mirrors with short lead time spheres and parabolas. All optics were over-sized to deliver a 10 arcsecond diameter beam to the scoring camera and guiding camera. A 5 arcsecond beam is sent to the polarimeter for flexible adjustment of system parameters.  All associated optical mounts were also commercial stock. Figure \ref{h85_optical_bench_install} shows the coud\'{e} experiment room and the curvature adaptive optics system during initial installation. The AEOS telescope primary and secondary mirrors (M1 and M2) create an f/200 beam.  This beam is folded around the azimuth and elevation axes then sent to the coud\'{e} rooms by an additional 5 flat mirrors (M3 to M7).  The beam is sent by the flat M7 to any of the six coud\'{e} rooms where the beam enters through a mechanical shutter in the wall seen in the top left of Figure \ref{h85_optical_bench_install}.   An end-to-end optical schematic with every element is shown at the end of this paper in Figure \ref{optical_path_layout}.

\begin{figure}
\centering
\includegraphics[width=0.60\textwidth]{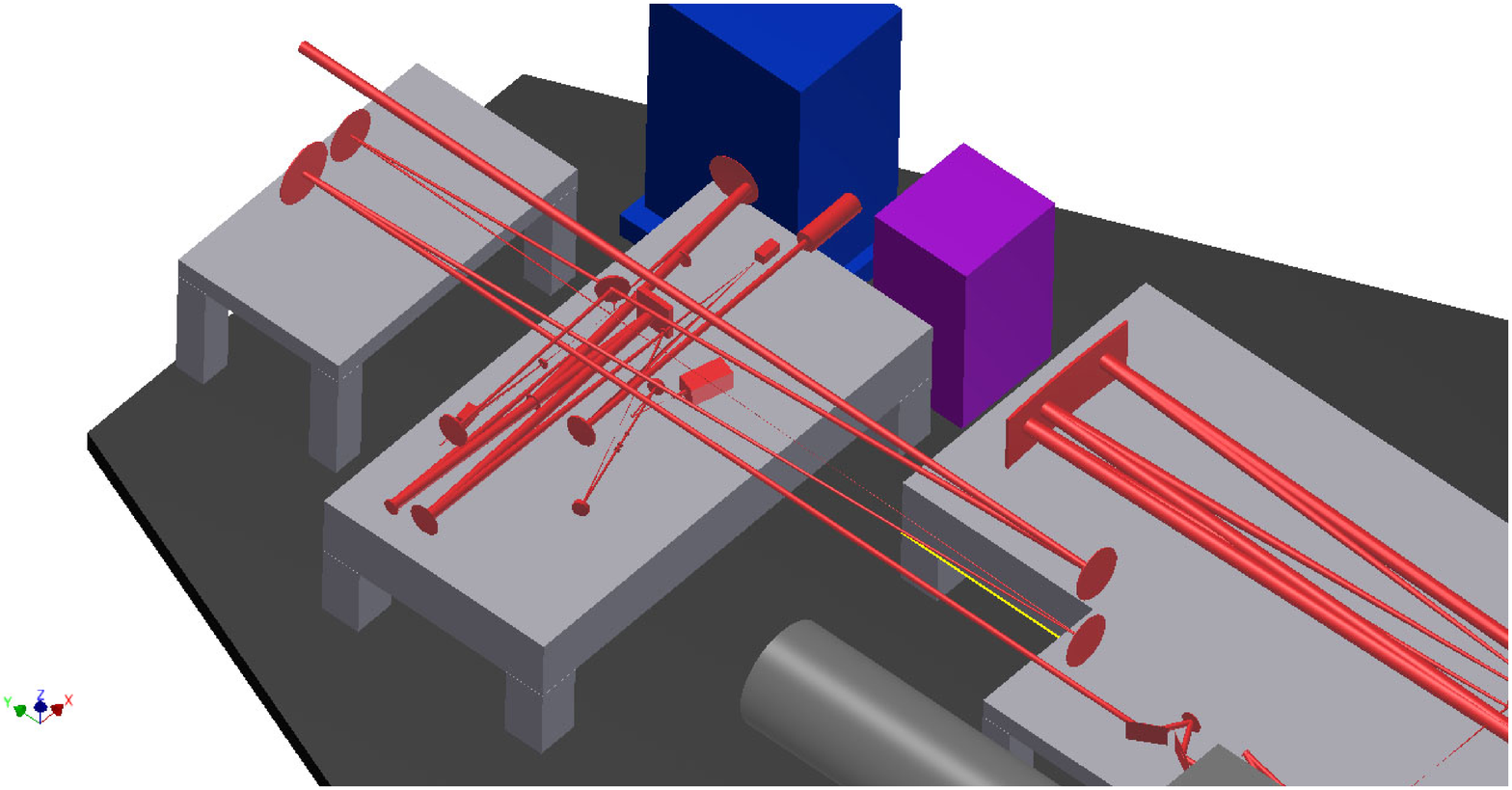}
\includegraphics[width=0.35\textwidth]{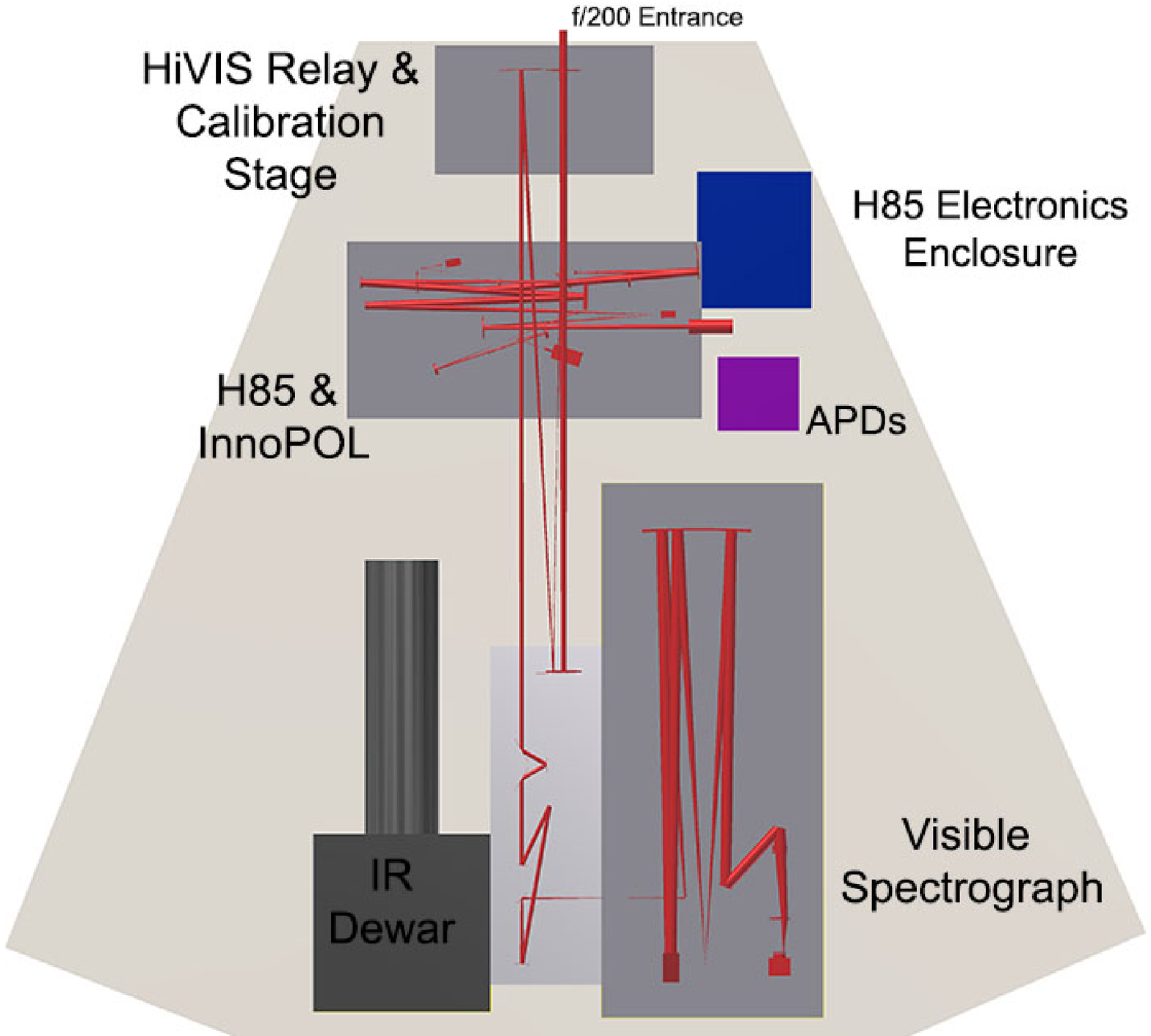}
\caption{The Zemax and 3D solid model for the H85, InnoPOL and HiVIS optical systems as built in the coud\'{e} room. See text for details.}
\label{zemax_model_room3}
\end{figure}

\begin{wrapfigure}{l}{0.7\textwidth}
\centering
\includegraphics[width=0.75\textwidth]{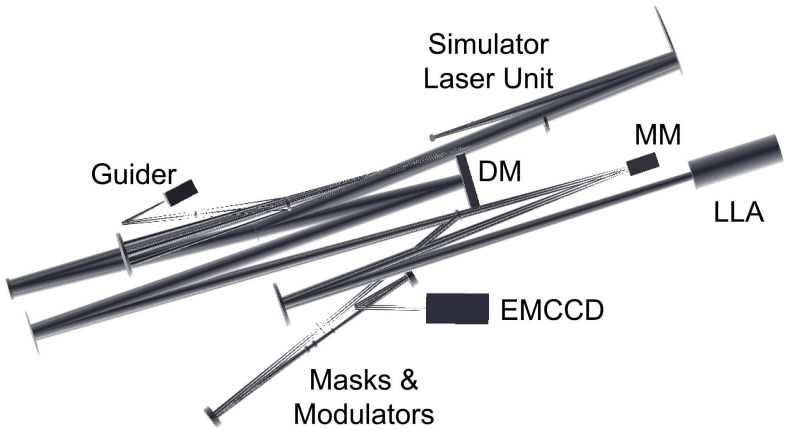}
\caption{The Zemax optical model for the adaptive optics bench including the guider camera, adaptive optics system and a 1-EMCCD test polarimeter. }
\label{zemax_model_ao_bench}
\end{wrapfigure}

The AEOS experiment room 3 houses the UH High resolution visible and infrared spectrograph (HiVIS) \cite{Thornton:2003bi, Hodapp:2000uo}. The new H-85 and InnoPOL instruments are built on a new optical bench sharing coud\'{e} experiment room 3 with the two HiVIS spectrographs. The HiVIS calibration injection port is shown in Figure \ref{h85_optical_bench_install} with a flat-field illumination screen in the beam path with the halogen lamps turned on. The coud\'{e} entrance shutter is immediately behind this calibration port.  The rest of the HiVIS optics are outside the image in Figure \ref{h85_optical_bench_install}. The H85 control electronics are contained in the large blue colored box of Figure \ref{h85_optical_bench_install}.  The black container in the top right of Figure \ref{h85_optical_bench_install} contains the 85 avalanche photo diodes (APDs) fed by optical fibers from the H85 wavefront sensor lenslet array. The lenslet array and membrane mirror are mounted just in front of the APDs in Figure \ref{h85_optical_bench_install}. 

Figure \ref{zemax_model_room3} shows the combined Zemax model for the H85 AO system, the InnoPOL polarimeter and the visible side of the HiVIS spectropolarimeter as built in could\'{e} room 3. The optical beam is shown in red.  The f/200 beam enters through the mechanical shutter in the wall at the far top left of Figure \ref{zemax_model_room3}.  

The incoming f/200 AEOS beam is brought to a convenient optical bench height by the first of the HiVIS relay mirrors (M8). This incoming beam is redirected to the new H-85 AO system by a removable flat pickoff mirror (MH0). The AO bench sits in the middle of the room as a self-contained optical unit.  The HiVIS spectrograph relay optics send the spectrograph beam across the room twice immediately above the H85 AO optical bench. This creates an opto-mechanical constraint on the AO system layout. We required fast switching between HiVIS and AO instruments which requires avoiding vignetting the HiVIS beam.

The H85 AO system has four major sub-systems that had to be designed with these constraints.  There is a curvature wavefront sensor, a deformable mirror to correct the wavefront, a guiding and acquisition camera and a telescope beam simulator for use in alignment, calibration and testing. 

The curvature adaptive optics system has a conventional optical layout.  A 50mm collimated beam is directed to the deformable mirror using three powered relay optics denoted MH1, MH2 and MH3.  This beam reflected off the DM is then focused on to the membrane mirror at f/40 with the MH4 mirror.  The beam reflected off the membrane mirror is then collimated by MH5 to deliver a 30mm diameter pupil image on the lenslet array (LLA). The lenslet array feeds avalanche photo diodes (APDs) for wavefront sensing. Table \ref{table_optical_path} shows the optical properties of the system optics without the polarimeter. 

\begin{wraptable}{l}{0.55\textwidth}
\caption{Optical Train - AEOS telecope, HiVIS fold, H85 relay optics, H85 AO system, InnoPOL polarimetric optics.}
\label{table_optical_path}
\begin{tabular}{cccl}
\hline
\hline
Optic 	& Conic 	& Fl 		& Optic			\\
Name	& Const	& in		& Description		\\
\hline
\hline
M1		& -1		& 215	& AEOS Primary			\\
M2		& -1.03	& -14.8	& AEOS Secondary, f200		\\
M3		& Flat	&		& AEOS M3				\\
M4		& Flat	&		& AEOS M4				\\
M5		& Flat	&		& AEOS M5				\\
M6		& Flat	&		& AEOS M6				\\
Window	& Flat	&		& AEOS coud\'{e} window 	\\
M7 		& Flat 	& 		& Final AEOS coud\'{e} mirror	\\ 
\hline
M8 		& Flat 	& 		& HiVIS fold to bench height	\\ 
\hline
MH0		& Flat	&		& Removable pickoff for H85	\\
MH1		& -1		& 48		& H85 Relay 1				\\
MH2		& 0		& 8		& H85 Relay 2				\\
Window	& Flat	&		& Guider Pickoff Window		\\
MH3		& 0		& 60		& H85 Relay 3				\\
\hline
DM		& Flat	&		& Deformable mirror	(DM)		\\
MH4		& -1		& 80		& f/40 focus to InnoPOL		\\
{\it Notch}	& Flat	&		& Dichroic Notch Filter (trans.)	\\
MM		& Flat	&		& Membrane mirror (MM)		\\
MH5		& -1		& 48		& Collimate to 30mm pupil	\\
LLA		& ---		&		& Lenslet Array	 WF sensor	\\
\hline
{\it Notch}	& Flat	&		& InnoPOL pickoff (before MM)	\\
I1		& -1		& 25.5	& InnoPOL collimator		\\
I2		& -1		& 25.5	& InnoPOL f/40 focus		\\
\hline
\hline
\end{tabular}
\end{wraptable}

A window in the relay optics diverts a small fraction of the broad-band incoming light to a guiding camera.  A dichroic just before the membrane mirror (MM) feeds a converging f/40 beam to the InnOPOL polarimeter.  This polarimeter includes field masks, occulting masks, pupil reimaging, liquid crystal modulators and rotating achromatic retarders.  

In order to accomplish this set of optical tasks given the constraints, we used 3 powered optics and one flat mirror in order to deliver a beam on to the optical bench while providing access for guiding and simulator beam injection.  The annotated Zemax model for the H85 optics is shown in Figure \ref{zemax_model_ao_bench}.

\subsection{AO Simulator Laser Beam}

H85 was built on AEOS with an internal simulator unit that injects a beam in to the AO system matching the key properties of the telescope beam.  A fiber-coupled laser diode is passed through several optics to create a collimated 50mm pupil image on to the DM matching the beam delivered by the H85 relay optics. This simulator has internal masks with the proper sized secondary obscuration ratio, and spider locations. There is an internal pupil location inside the simulator where rotating phase masks and polarization calibration optics can be inserted for system characterization.  We have single mode fiber couple laser diodes and broadband fiber coupled sources to simulate various telescope beams.

In the most typical simulator configuration, we have a single mode fiber combiner that combines two independent laser sources (635nm and 660nm) in to one single-mode output. We currently have in place a Thor labs reflective notch filter as a dichroic for use with the scoring camera and polarimeter. This reflective notch filter splits $\sim$10nm of bandpass to the scoring camera or polarimeter while sending the remaining light to the wavefront sensor.  This dichroic has transmission spectra measured with our Varian Cary spectrophotometer.  The bandpass is a smooth and reasonable function of tilt angle as expected for these type of dielectric stack reflective notch filters.  We also took special care to select the notch filters which will not introduce non-common path wavefront errors.  The beam footprint on the notch filter is less than 3 mm.  The major uncertainty is the wavefront error introduced by the dielectric coating stack in the beam reflected to the science camera. To overcome this potential error source, we measured the reflected wavefront flatness with our Zygo interferometer.  We found that the major wavefront error term was dominated by the focus term. Our filters all showed a few waves of focus but were all better than $\lambda$/10 to $\lambda$/20 when focus is removed.  This means the science channel will be at a slightly different f/ number than the beam to the wavefront sensor. However this has negligible consequence for both calibration and operations.

\subsection{Tip/Tilt control systems}

The deformable mirror cell mount itself is an active tip tilt control system.  Two Physik Instrumente 239 actuators with 180 micron stroke are used in a kinematic configuration to tip and tilt the cell holding the deformable mirror.  The wavefront sensor control system computes the average wavefront tilt and offloads this signal to the DM cell.  This control system has a nominal bandwidth limit of roughly 100Hz.  The simulator laser system is used to derive tip/tilt control system interaction matrices (IMATs) which are used to minimize the tip/tilt corrections required on the DM itself.

\subsection{Acquisition \& Guiding}

Acquisition and tracking of targets requires a reasonably fast, stable and sensitive camera capable of driving the telescope beam to the center of the wavefront sensor.  The wavefront sensor has a very small field of view making the guiding system stability and flexibility critical to overall instrument function. The H85 relay optics provide two intermediate beams at convenient heights off the optical bench.  A transmissive pickoff window is placed in the beam to divert a small fraction of broad-band light to the guiding system.  

\begin{wrapfigure}{r}{0.5\textwidth}
\centering\includegraphics[width=0.5\textwidth]{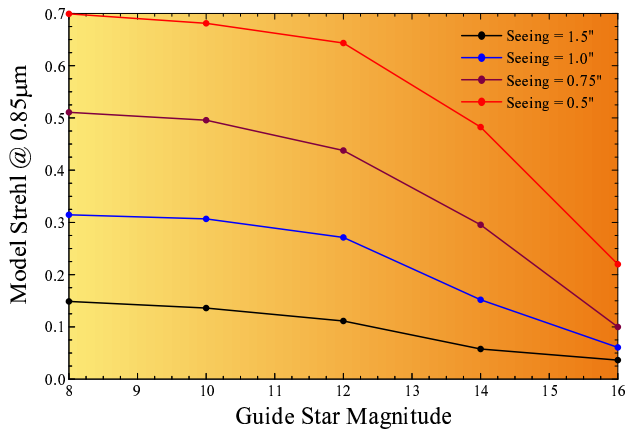}
\caption{The predicted Strehl ratio at 850nm delivered by H85 to the scoring camera f/40 focus. }
\label{h85_strehl}
\end{wrapfigure}

This beam is folded and focused on to a simple SBIG ST-402ME guiding camera.  Commercial software (Maxim DL) is used to derive tip/tilt corrections and output the offsets.  We have developed in-house software to take these guiding offsets and send appropriate centering commands to the AEOS mount gimbal control system.

\subsection{Predicted AO Performance}

End-to-end simulations of the curvature AO system performance have been created in Matlab. Over the creation of the various UH 36 and 85 element curvature AO systems the model performance has been substantially validated against the H-85 and NICI instruments.  These programs take in atmospheric wind speed, turbulence strength ($r_0$), guide star brightness, AO system IMATs, gain and control system settings. The simulated image output can be easily analyzed for optical performance, delivered Strehl ratio and other parameters. Figure \ref{h85_strehl} shows the model H85 delivered Strehl ratio under varying seeing conditions at an observing wavelength of 850nm. The guide star brightness and wavefront sensor performance was derived using the predicted instrument throughput as installed at AEOS in it's present configuration.  In 2008-2010, a test version of the H85 system was built on AEOS working at f/56.  With this test system, we were able to verify our throughput models to the H85 wavefront sensor and validate the predicted flux levels delivered to the coud\'{e} room.

\begin{figure}
\centering
\includegraphics[width=0.99\textwidth, height=0.44\textwidth, angle=0]{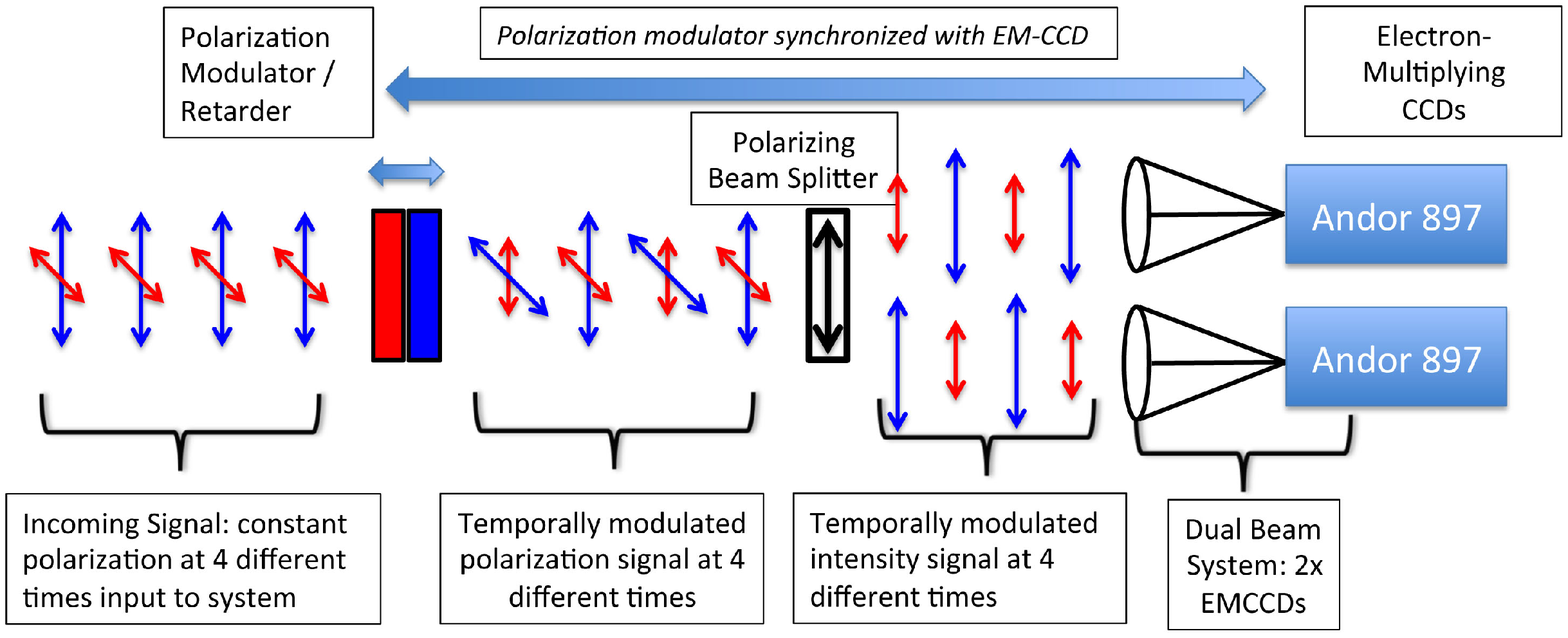}
\caption{The conceptual model for our polarimeter. The incoming light is unpolarized with orthogonal incident polarizations represented as different colors. Each set of arrows corresponds to 4 different times separated by half the modulation frequency. The modulator acts to re-orient this input polarization as a function of time shown in the middle. The analyzer sends different modulated components in to two different beams. Each beam is recorded on it's own EMCCD with each frame readout synchronized with the modulator. }
\label{polarimeter_schematic}
\end{figure}

\subsection{Instrumental Approach: Polarimetric modulation and EMCCDs} 

	Modern frame-transfer electron-multiplying detectors (EMCCDs) are becoming quite close to this ideal. By making observations faster than the characteristic atmospheric time scale, we freeze speckles from turbulence and do our differential photometry with a pseudo-static optical system. We average many pairs of modulated images to improve statistics. CCD frame-transfer storage regions allow for the readout of one image to occur while the subsequent image is being exposed, giving a 100\% duty-cycle with no time lost for readout. The newest generations of EMCCDs can continuously read 128x128 pixel frames at speeds over 500Hz allowing $<$2ms polarization modulation. Polarimetric modulation is also now possible with a range of components at high speeds.  Several commercial options exist using liquid crystal technologies that switch faster than 0.2ms and have easily tunable optical properties. A schematic overview of this modulation plus fast-frame-rate readout polarimeter is shown in Figure \ref{polarimeter_schematic}.

There are many examples of these technologies in solar and night-time applications. An early approach in solar imaging polarimetry was the Zurich Imaging Polarimeter (ZIMPOL)  \cite{Stenflo:2007wq,Stenflo:2003io,Keller:1994ww}.  Masked arrays with bi-directional charge clocking were used at the focal plane and fast-switching  (1kHz)  modulation was performed ferro-electric liquid crystals (FLCs). Here the photoelectric charge from each FLC-modulated image is accumulated in masked storage regions until the pixel wells are full. 

The only night-time instruments presently planned or on-sky with this technology for AO-assisted imaging polarimetry is the VLT SPHERE \cite{Roelfsema:2010ca,Gisler:2004ck,Thalmann:2008gi} and the EPOL concept designed for the ELT \cite{Keller:2010ig}. We note that the Gemini Planet Imager (GPI) and HiCIAO on Subaru have AO-assisted dual-beam polarimeters but not of the sophistication described here \cite{Macintosh:2008dna,Perrin:2010dw,Suzuki:2010by,Tamura:2006hr}. EMCCDs and FLCs have been used without AO correction in ExPo \cite{Rodenhuis:2008ji, Canovas:2011ed, Rodenhuis:2012du, Canovas:2012ib,Min:2012ha,Jeffers:2012ki}. The technology we propose here has not been used in the US on night-time telescopes.

\subsection{Polarimetric Imager Optical Design}

\begin{wrapfigure}{r}{0.65\textwidth}
\centering
\includegraphics[width=0.64\textwidth, angle=0]{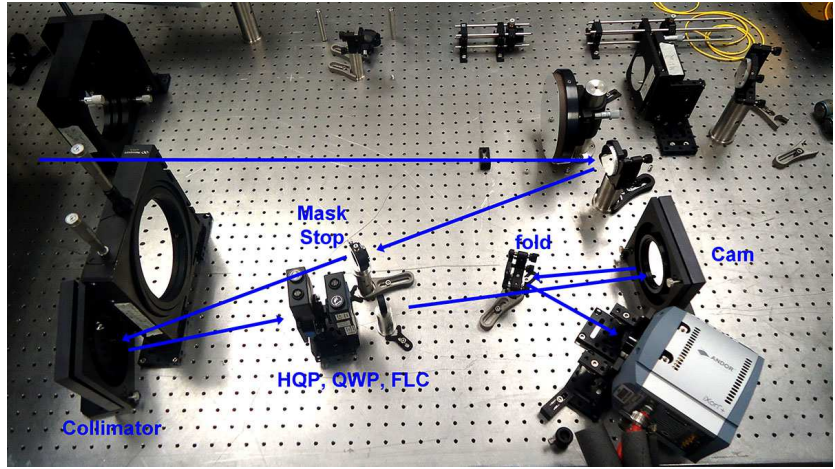}
\caption{The optical layout of the polarimeter optics with a single channel test EMCCD. See text for details.}
\label{polarimeter_image}
\end{wrapfigure}

The AO system delivers a compensated f/40 focus at a convenient bench height just after reflection off the reflective notch filter as entrance to the polarimeter. The polarimeter beam is collimated to produce a beam suitable for polarimetric modulation techniques. A nominal 11mm pupil image is formed in a beam that provides enough space for both liquid crystal modulators and rotating crystalline or polymer-based retarders.  For the single-EMCCD configuration, his collimated beam is then focused to an EMCCD by another powered mirror.  A small fold mirror directs the converging beam on to an EMCCD for convenient mechanical placement. A Savart plate is placed just in front of the EMCCD to provide a dual-beam polarimetric image.  Initial testing was done with a single Andor iXon EM$^+$ EMCCD camera in 2012. 

Figure \ref{polarimeter_image} shows an early test of this optical layout using simple flat mirrors in place of the dichroic and deformable mirror. The beam comes to the f/40 focus where field stop masks are located. During the design phase of the polarimeter, we included the option for focal plane masks and transmissive occulting masks at the AO-corrected f/40 focus. We have procured many photo-lithographic ND filter masks where chrome is deposited on a glass substrate with ND3 and ND5 densities in spot diameters ranging from 50 to 500 microns. 

\begin{wrapfigure}{r}{0.55\textwidth}
\centering
\includegraphics[width=0.55\textwidth]{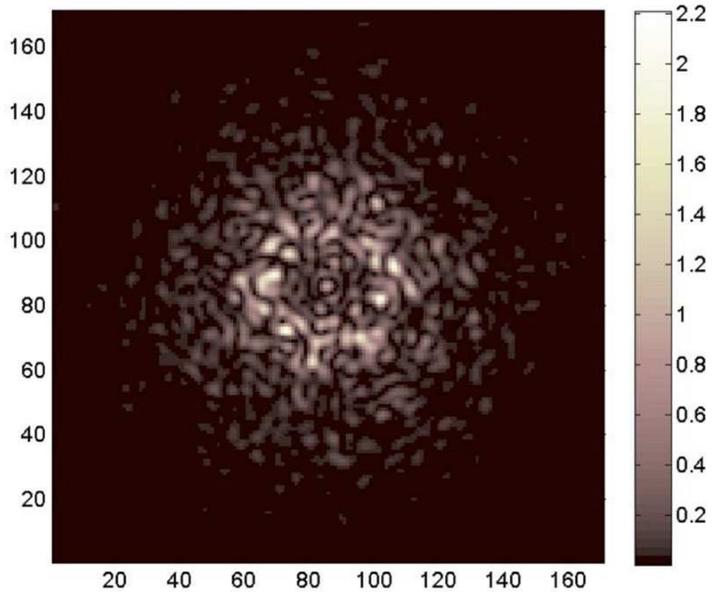}
\caption{A simulated 2 millisecond exposure at the AO corrected InnoPOL focus using an ND3 occulting spot at the f/40 focus and coronographic Lyot stop at the liquid crystal filter pupil image. Sampling was set to 24 milliarcseconds per pixel.}
\label{h85_simultated_intensity}
\end{wrapfigure}

The collimating mirror sends the beam to the modulators.  Nominally we will be using Ferro-electric type liquid crystals (FLCs) to perform modulation at speeds up to 5kHz. These FLCs are placed at the pupil image.  Additionally, coronagraphic masks can be placed adjacent to the FLCs.  Slower modulation of the telescope beam is accomplished with rotating retarders placed in computer-controlled stages.  In a traditional configuration a combination of a half-wave retarder plate (HWP) and a quarter-wave retarder plate (QWP) are placed in front of the FLC to accomplish a standard Stokes definition modulation scheme.  These can be seen in the center of Figure \ref{polarimeter_image}.

\subsection{Predicted Polarimetric Camera Performance and Key Trade Offs}

From the AO performance models, we create simulated delivered focal plane intensities at high temporal and spatial resolution. A single 2ms exposure simulation is shown in Figure \ref{h85_simultated_intensity}.  We have included a Lyot-type coronagraphic mask at the InnoPOL pupil image and an associated 100 micron ND3 occulting spot masking the focal plane. 

The residual speckles dominate the $\sim$3 arc second diameter focal plane of Figure \ref{h85_simultated_intensity}. With the AO correction, the image underlying the speckles has high spatial resolution. However the dominant source of noise will be the speckles.   In order to accomplish effective polarimetric imaging, both the spatial and temporal properties of the speckles must be suppressed to levels at or below the statistical shot noise and other limiting EMCCD detector noise sources. 

Cost effective commercial EMCCDs have fixed readout rates with noise levels increasing with readout speed.  Read noise can be compensated somewhat with increased gain but at the sacrifice of dynamic range in addition to an effective reduction in quantum efficiency from the gain register induced noise factor.  Polarimetric modulation schemes are driven to fast speeds in order to correlate and suppress speckle brightness and focal plane location. 

Instrument performance requirements directly opposed to high speed modulation is the field of view (FOV), spatial sampling especially when considering scenes with dynamic range.  The dynamic range requirement is absolutely critical in applications with faint objects next to bright objects such as detecting exoplanets, circumstellar disks and SSA are search applications.   

\begin{figure}
\centering
\includegraphics[width=0.45\textwidth]{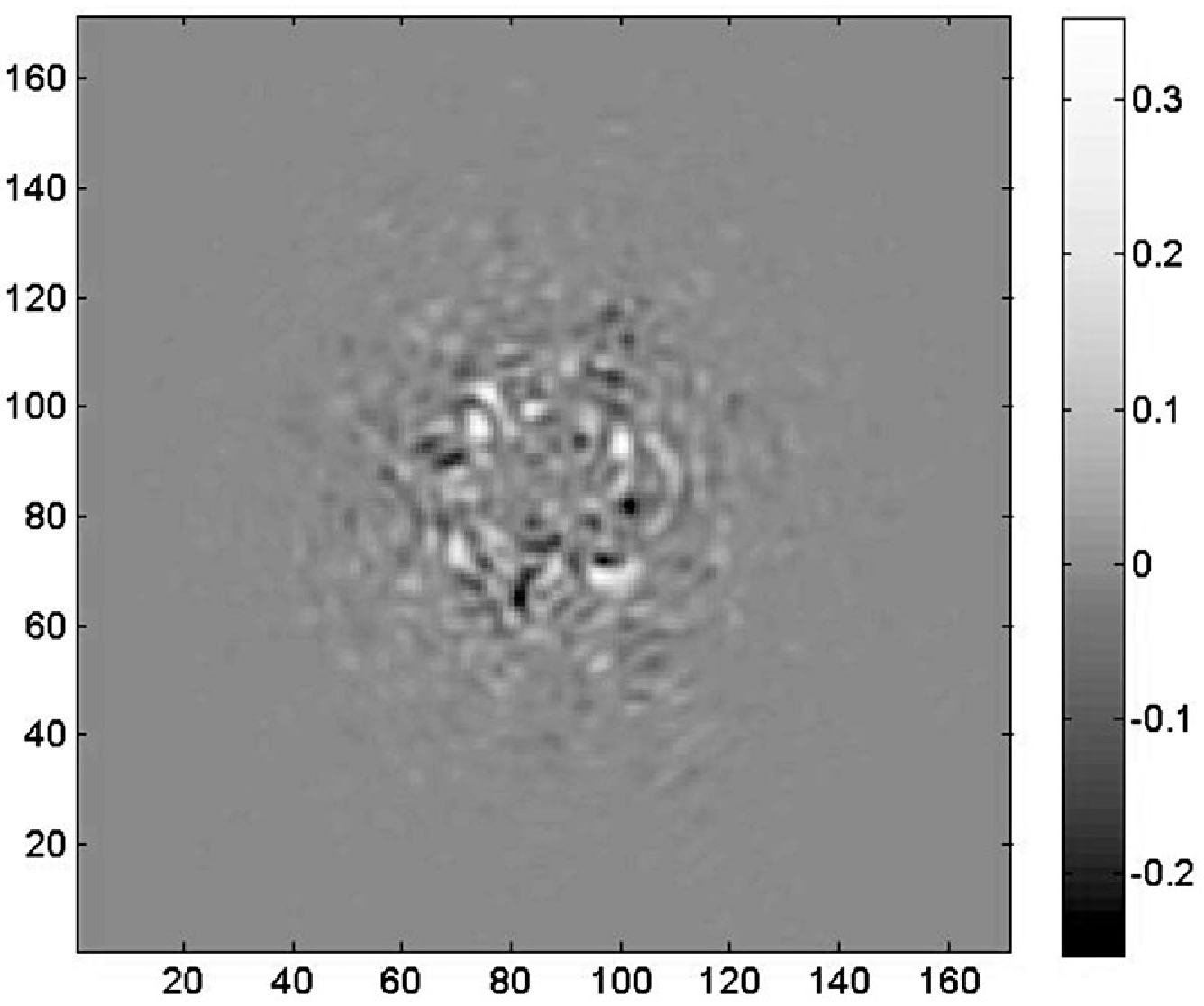}
\includegraphics[width=0.50\textwidth]{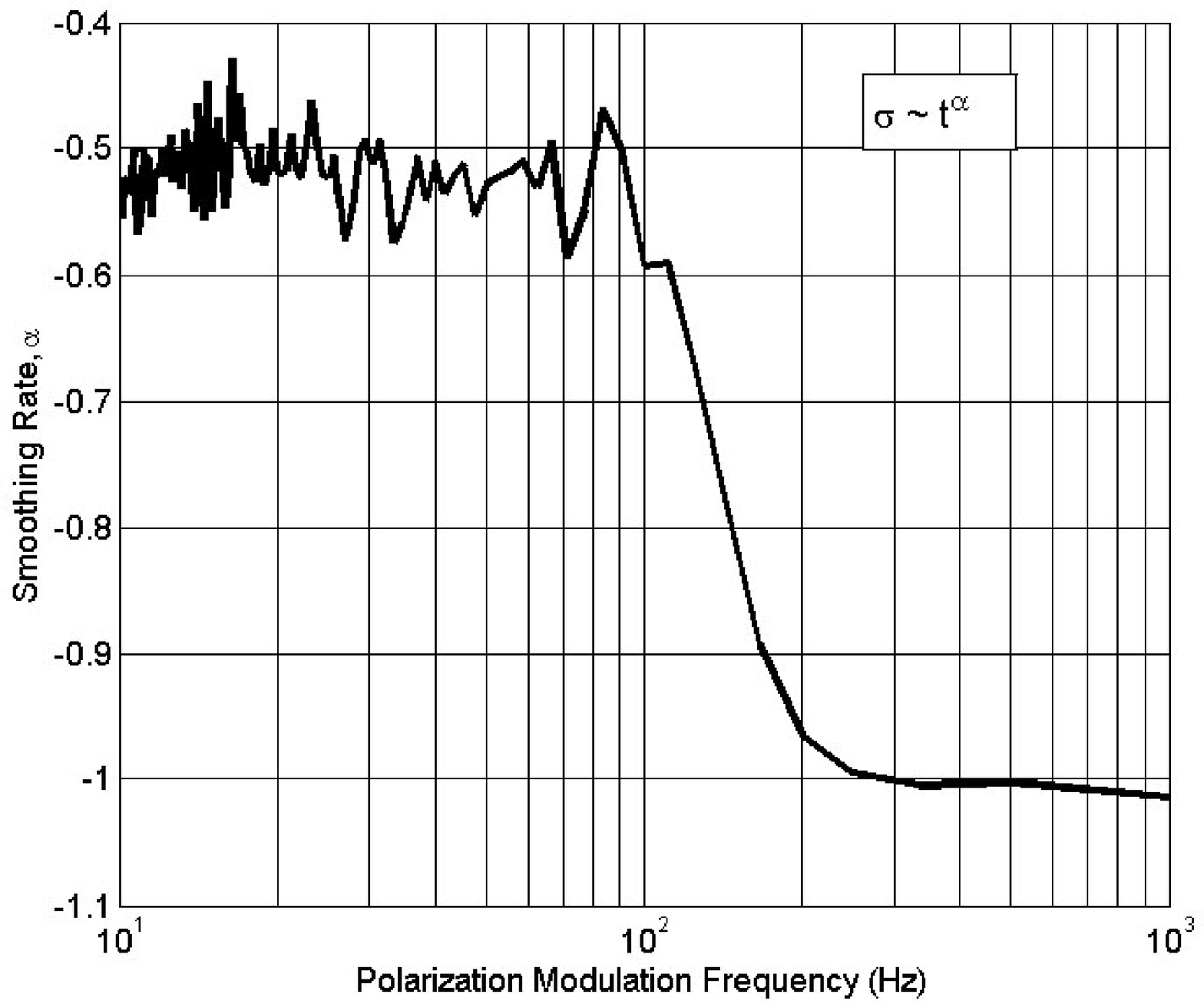}
\caption{a) The predicted differences between two successive 2ms exposures computed as a ratio $(I1 - I2)/(I1 + I2)$. This mimics spurious polarization errors caused by speckle evolution  b) The exponent of the focal plane rms variation with time in the focal plane computed for varying frame rates. Statistically independent variations smooth like $\sqrt{n}$ (-0.5) whereas correlated variations smooth like 1/n. }
\label{speckle_smoothing_simulations}
\end{figure}

Using our high time resolution intensity distributions, we can compute the simulated residual speckle changes both in brightness and location as functions of EMCCD frame rate.  Figure \ref{speckle_smoothing_simulations} shows an example of residual speckle errors computed as a normalized ratios $(I1 - I2)/(I1 + I2)$ of subsequent 2ms exposures (500Hz frame rate).  Each normalized ratio is computed pixel by pixel for each image as $(I1 - I2)/(I1 + I2)$ as would typically be performed in a simple polarimetric data reduction calculation. 

In any dual-beam polarimetric system with multiple redundant modulated images, there are several ways to compute a polarization signal given a sequence of temporally modulated dual-beam images. Different methods are sensitive to different systematic errors and depend on the modulation scheme. For instance, one can do normalized difference ratios $(I1 - I2)/(I1 + I2)$ using different pixels imaged at the same time or using identical pixels with temporal modulation. For the sake of simplicity, we assume only temporal modulation and treat each beam independently for this analysis.

A statistical analysis of the resulting residual speckle-induced polarization noise images $(I1 - I2)/(I1 + I2)$ shows a substantial change in noise behavior as modulation speed increases. An example of a speckle image computed from our system performance models is shown in the left panel of Figure \ref{speckle_smoothing_simulations}.  To quantify this polarization noise, the root-mean-square (RMS) error of the individual polarization frames $(I1 - I2)/(I1 + I2)$ is computed. To calculate the polarization noise at different modulation rates, the polarization speckle-induced noise images are computed using a single model run with images computed at a large range of image exposure times. Polarization noise is computed for each simulated EMCCD frame rate. This RMS error of the resulting polarization speckle noise is a good measure of the background noise level caused by speckle evolution for narrow FOV simulations where speckle nose dominates and scene dynamic range variations are relatively small.  The decrease in noise should follow the statistical smoothing of independent atmospheric phase screen realizations in the slow limit ($\sqrt(n)$).  The decrease in polarization noise is computed between each successive data set at varying EMCCD frame rates. The exponent for the power-law decrease in polarization noise is computed between each successive simulation.  The right panel of Figure \ref{speckle_smoothing_simulations} shows RMS error of the polarization images as EMCCD frame rates are increased from 10 Hz to 1000 Hz.  

For independent realizations of a statistical process at very slow EMCCD frame rates, smoothing of an image would follow a $\sqrt{t}$ type process. The corresponding speckle noise would be expected to smooth as independent realizations are averaged.  However, at modulation frequencies above 200Hz, the simulations show that speckles become correlated in successive EMCCD frames and the speckle induced polarization noise starts to smooth at faster rates of $\frac{1}{time}$. At high modulation rates, speckles are essentially frozen in both brightness and focal plane location. Speckles will effectively difference in polarization calculations leaving only shot noise variations. The shot noise in these polarization frames goes like $\sqrt{N} = \sqrt{I1 + I2}$ so bright speckles correspond to regions of low noise. Furthermore, polarization is computed on a pixel by pixel basis making it independent of individual pixel gain variations (flat fielding). 

The result that atmospheric effects induce speckle brightness and location changes occur on fast timescales is straightforward. Given the Fried length $r_0$, the telescope aperture $D$, and the wind speed $v$, there are two relevant time scales we can build. The first,  $\tau_0 = r_0/v$, the time it takes a coherent phase patch to move its own length and $\tau_s = D/v$ the time it takes a phase screen to move across the aperture. The timescale 0.5$\tau_s$ is associated with a turnover of the whole speckle pattern whereas Roddier et al. \cite{Roddier:1982du} have argued that $\tau_b = 0.3\tau_0$ is the relevant time scale for speckle boiling. As individual patches of size $r_0$ enter and exit the telescope aperture, relative brightness changes in speckles is expected. Given the 3.7m AEOS telescope, a Fried length of 15cm (0.7 arcsecond seeing at 500nm ) and a wind speed of 10m/s we find $\tau_s$ =185ms and $\tau_b$ =5ms. Our simulations presented in Figure \ref{speckle_smoothing_simulations} are entirely consistent with RoddierÕs analysis \cite{Roddier:1982du}.  Similar results have been obtained by others studying speckle suppression in AO systems and associated coronagraphs. \cite{2004ApJ...610L..69B, 2004ApJ...612L..85A, Codona:2013iw, Ren:2012ec, Fitzgerald:2006wv, 1991A&A...243..553V, 2010PASP..122...71G, 2004ApJ...615..562G} In our system, the DM and even a single atmospheric turbulence layer create two time dependent phase screens that evolve independently and interact coherently. Our AO system models assume a single turbulence layer and a single wind speed yet still produce correlations and speckle suppression with modulation speeds of order 5ms.

\begin{wrapfigure}{l}{0.40\textwidth}
\centering
\includegraphics[width=0.39\textwidth, angle=90]{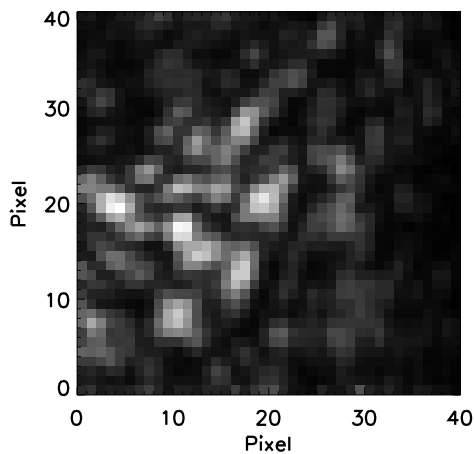}
\caption{A single 0.5ms image recorded with an EMCCD camera at the f/40 focus. The DM and notch filters were replaced with flat mirrors for high flux speckle evolution testing.}
\label{speckle_image_emccd}
\end{wrapfigure}

The theoretical timescales for speckle evolution depends critically on the assumed instrument. For example, Macintosh et al \cite{Macintosh:2005ih} outline two separate temporal regimes: one where diffraction effects are dominant and so-called first order speckles modulate rapidly as atmospheric phase variations evolve with the $\frac{r_0}{v}$ crossing time or faster. The second regime is coronagraphic with apodized pupil masks where speckle lifetimes are dominated by the phase clearing timescale of $\frac{D}{v}$. In the Macintosh et al \cite{Macintosh:2005ih} analysis, the so called first-order speckles are safely ignored as they are strongly suppressed in the high strehl AO assisted coronagraphic use case using apodized pupil masks. The analysis also shows fast intensity modulation at much higher amplitudes without and apodized coronagraph mask.  For dual-beam imaging polarimetry in our use case the first order speckles dominate the PSF halo outside the first few airy rings and drive the noise floor limitations in polarimetric images. Our H85 system models with and without coronagraphic masks clearly show these first order speckles as the primary noise source outside the image core. For dual-beam imaging polarimetry, speeding up the polarimetric modulation substantially reduces the frame-to-frame speckle noise allowing for less stringent instrument requirements on other system components to achieve polarimetric precision design goals. Our initial EMCCD tests and many H85 performance simulations show these first order effects clearly dominating the polarimetric image halos as the limiting source of polarimetric noise.

\section{INNOPOL EMCCD \& LIQUID CRYSTAL POLARIMETER}

The EMCCD polarimeter was designed to be a flexible test bench for a dual beam system. In 2012, we purchased two Andor iXon Ultra 897 EMCCDs for use in a dual-channel polarimeter.  Associated control electronics for liquid crystal drive signals, camera synchronization and triggering were purchased from Agilent and Tegam. Using ethernet-controllable function generators allows us to drive several liquid crystal types including ferro-electric and nematic. 

We purchased polarimetric components to allow dual-beam polarimetry in configurations using either one and two EMCCDs. Wire grid polarizers make effective polarizing beam splitters. The reflected beam when created by  a cover-glass free wire grid polarizer shows good image quality and high degree of polarization in laboratory testing. Using both the reflected and transmitted beams creates two independent imaging channels.  A 2-camera system provides the opportunity to double the pixel readout rate and possibly reduce systematic errors but with increased cost, system complexity.  A calcite Savart plate was also purchased to allow testing in a dual-beam polarimeter with only a single EMCCD detector.  This simplifies system complexity and operation but with a 2x reduction in the pixel readout rate capability. Given the fold mirror and open polarimeter optical bench space shown in Figure \ref{polarimeter_image}, we can insert several modulator options as well as change camera mirror focal lengths and effective sampling.  Expansion of the modulators to include achromatic liquid crystal designs is easily feasible following common techniques in solar and stellar applications. \cite{Gisler:2003hy, deWijn:2010fh,Tomczyk:2010wt}

The InnoPOL optics first collimate the f/40 AO corrected beam. This creates an 11mm diameter pupil image as well as provides a collimated beam for slow and fast polarimetric modulation.  Space is also created for corona graphic masks.  In the wire-grid polarizer 2-channel setup, the wire grid then splits the two beams and two independent images are formed with two independent powered optics.  In the single-channel system, a powered optic focuses the beam that is folded on to a single EMCCD. A Savart plate placed just in front of the converging beam creates a dual-beam polarimeter.  A complete layout is shown at the end of this document in Figure \ref{optical_path_layout}.

\subsection{Speckle Induced Noise Measurements - Atmosphere \& Telescope Jitter}

An early test we performed on sky was an assessment of the speckle evolution combined with any telescope or instrument vibration by using high frame rate image sequences on bright stars at varying zenith angles. The EMCCD polarimeter as installed on the AEOS coud\'{e} 3 optical bench in the standard single-channel configuration.  An Andor iXon Ultra 897 was installed at the polarimeter focus.  No modulators or other active components were mounted in the optical path.  The AO system interaction matrices were recorded with the simulator laser and deformable mirror was flattened.  Bright stars were then acquired and focused on the scoring camera at f/40.  A sample 0.5 millisecond exposure taken with an 800nm $\pm$10nm bandpass filter is shown in Figure \ref{polarimeter_image}.  

\begin{figure}
\centering\includegraphics[width=0.36\textwidth, angle=90]{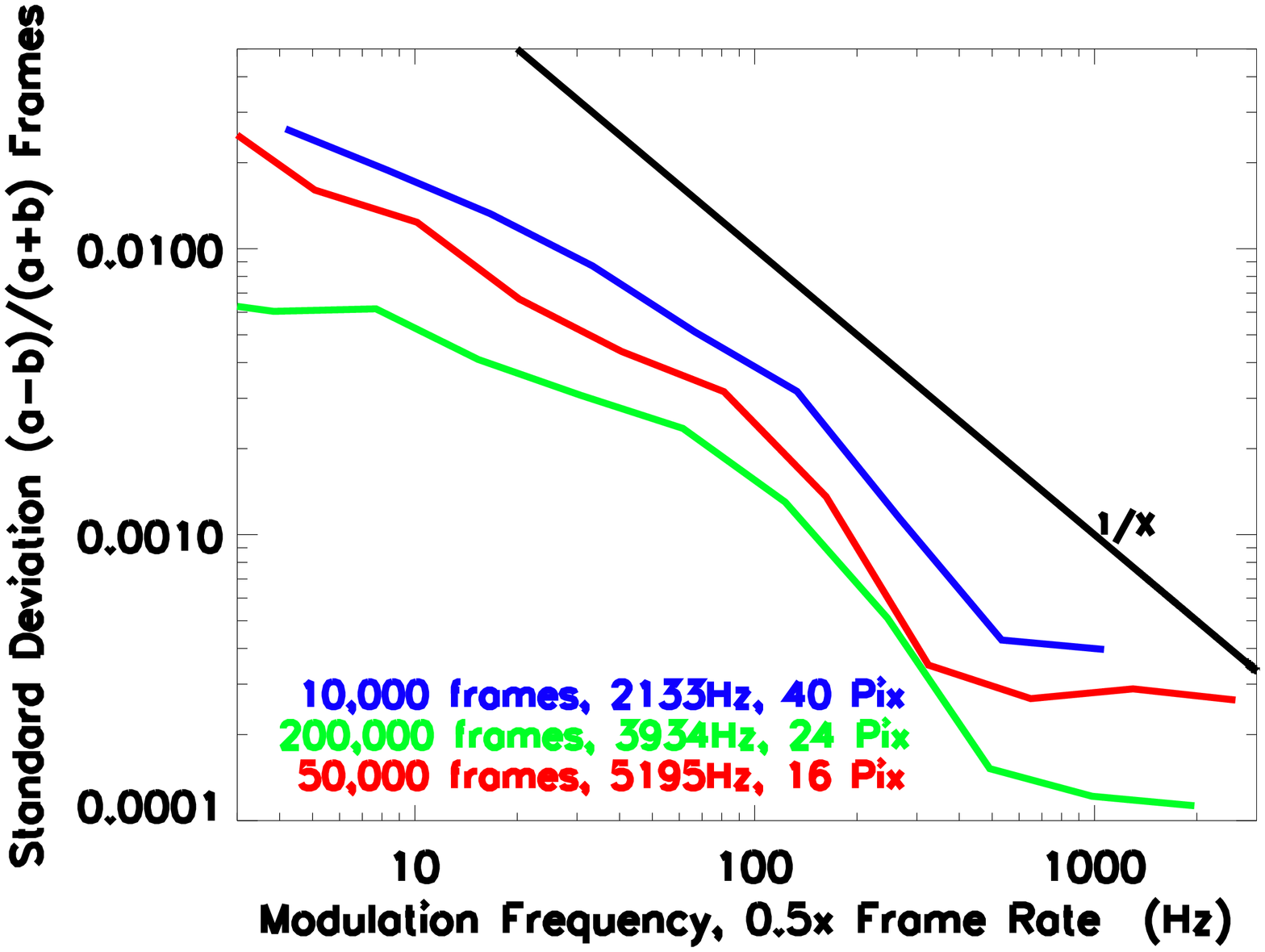}
\centering\includegraphics[width=0.36\textwidth, angle=90]{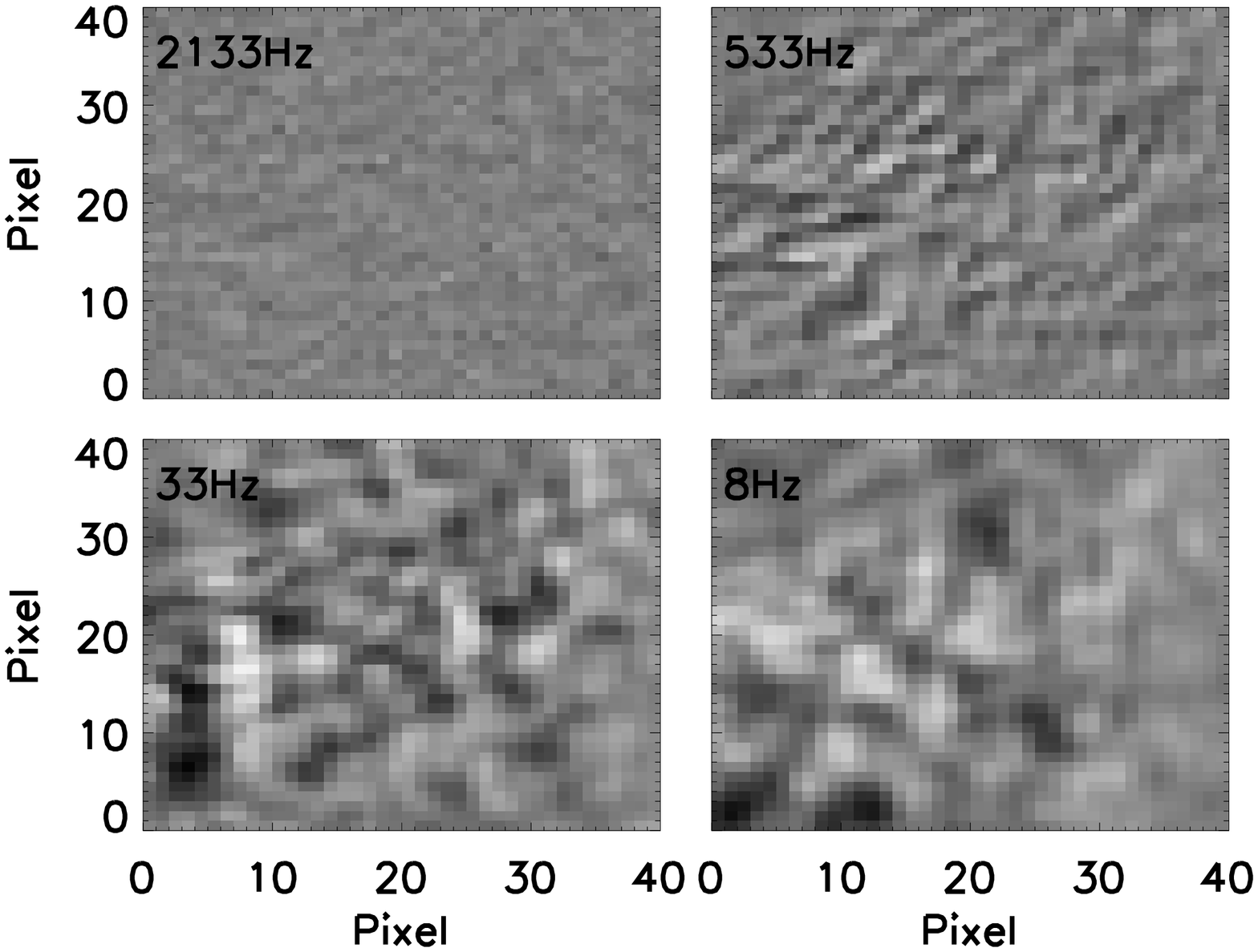}
\caption{a) The measured rms variation across an entire polarization image (a-b)/(a+b) computed after temporal binning to simulate varying EMCCD frame rates.  b) Snapshots of the corresponding polarization images at different modulation rates to show speckle induced polarization noise morphology. Each snapshot is on an identical grey scale.}
\label{emccd_speckles}
\end{figure}

During this test observing run, several independent acquisitions were obtained with different EMCCD frame rates on different stars at different zenith angles.  Figure \ref{emccd_speckles} shows three data sets obtained in integrations lasting a few minutes. The Andor SOLIS software package was used in the fast {\it crop sensor} and frame transfer mode with associated field stop masks installed to minimize light leakage. Acquisitions were collected with 2.1kHz frame rate with 40 by 40 pixel images, 3.9kHz frame rate with 24 by 24 pixel images and finally 5.2kHz frame rate with 16 by 16 pixel images. For each frame rate setting, the EMCCD gain was adjusted to maintain detected count levels roughly 20\% below the upper limits of the analog-to-digital converter. The electron multiplying gains were in the range of 50 to 100 with the appropriate pre-amplifier gain setting.  For the Andor Ultra 897 EMCCDs, the ccd pixel full wells are 125,000 electrons and the gain register pixel full wells are 800,000 electrons.  The effective read noise is in the $\sim$100 electron equivalent range with photo-electron levels amplified close to the 800,000 per pixel limit at the readout amplifier.  Photon statistical noise dominates the focal plane images.

To measure the speckle-induced polarimetric noise as a function of EMCCD frame rates, these data sets were processed by performing a simple polarimetric calculations after temporal averaging of sequential images in varying blocks.  Images were co-added temporally using in successive powers of 2 to provide image pairs at successively slower effective frame rates.  These temporally averaged image sequences can then be processed using a normalized difference calculation $(I1 - I2)/(I1 + I2)$ as would be standard in a polarimetric analysis.

Without any AO correction, speckles and high flux levels were detected with minimal dynamic range variations across the focal plane.  The left panel of Figure \ref{emccd_speckles} shows how the RMS speckle noise varies with changing temporal binning to simulate varying image frame rates. This allows us to compare our observations with the simulations of Figure \ref{speckle_smoothing_simulations}. Each colored curve shows the residual polarimetric noise as frame rate varies. Each curve is computed with an identical number of frames.  Statistical noise (shot noise) and read noise are the same for every effective frame rate leaving speckle-induced noise as the dominant error source. With effective frame rates of 10-50Hz, noise levels are above 0.1\% for each data set.  Beginning around 100Hz, the residual speckle-induced noise begins to drop substantially showing speckle correlations.  At effective frame rates in the 200Hz to 400Hz range, noise limits approach 0.01\% with other limiting noise sources showing at the highest frame rates.  The slowest effective frame rates (2Hz) show noise levels of roughly 1\% or worse.  For slow frame rates of 1-50Hz the noise roughly decreases as $\sqrt{n}$ 

The right panel of Figure \ref{emccd_speckles} shows sample polarization calculations $(I1 - I2)/(I1 + I2)$ at 4 different effective frame rates. At slow frame rates, the differences are essentially independent realizations of a random statistical process which smooths as $\sqrt{n}$.  At the fastest frame rates, only small speckle changes and residual readout noise are seen.

\begin{figure}
\centering
\includegraphics[width=0.38\textwidth, angle=90]{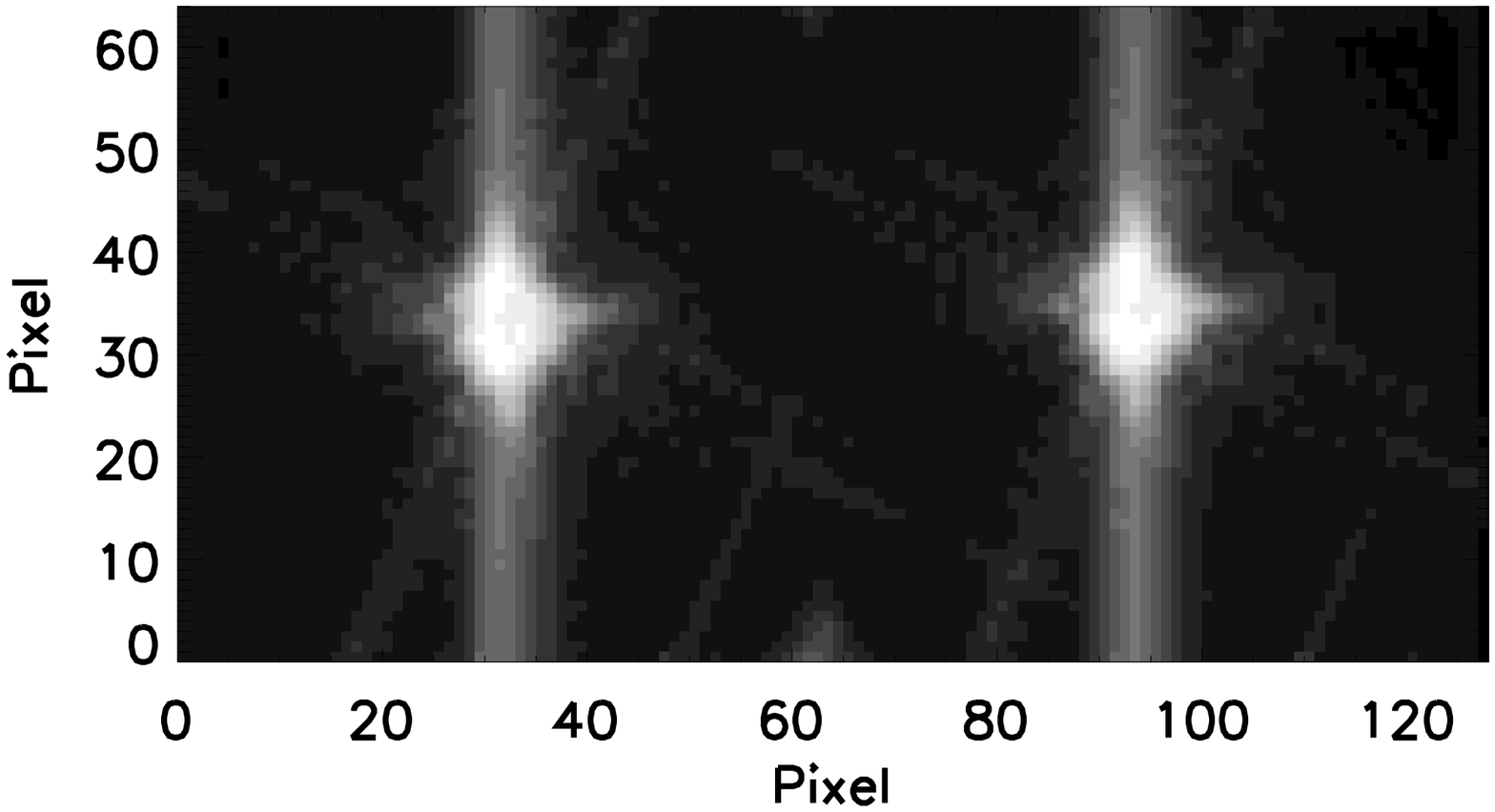}
\includegraphics[width=0.33\textwidth, angle=90]{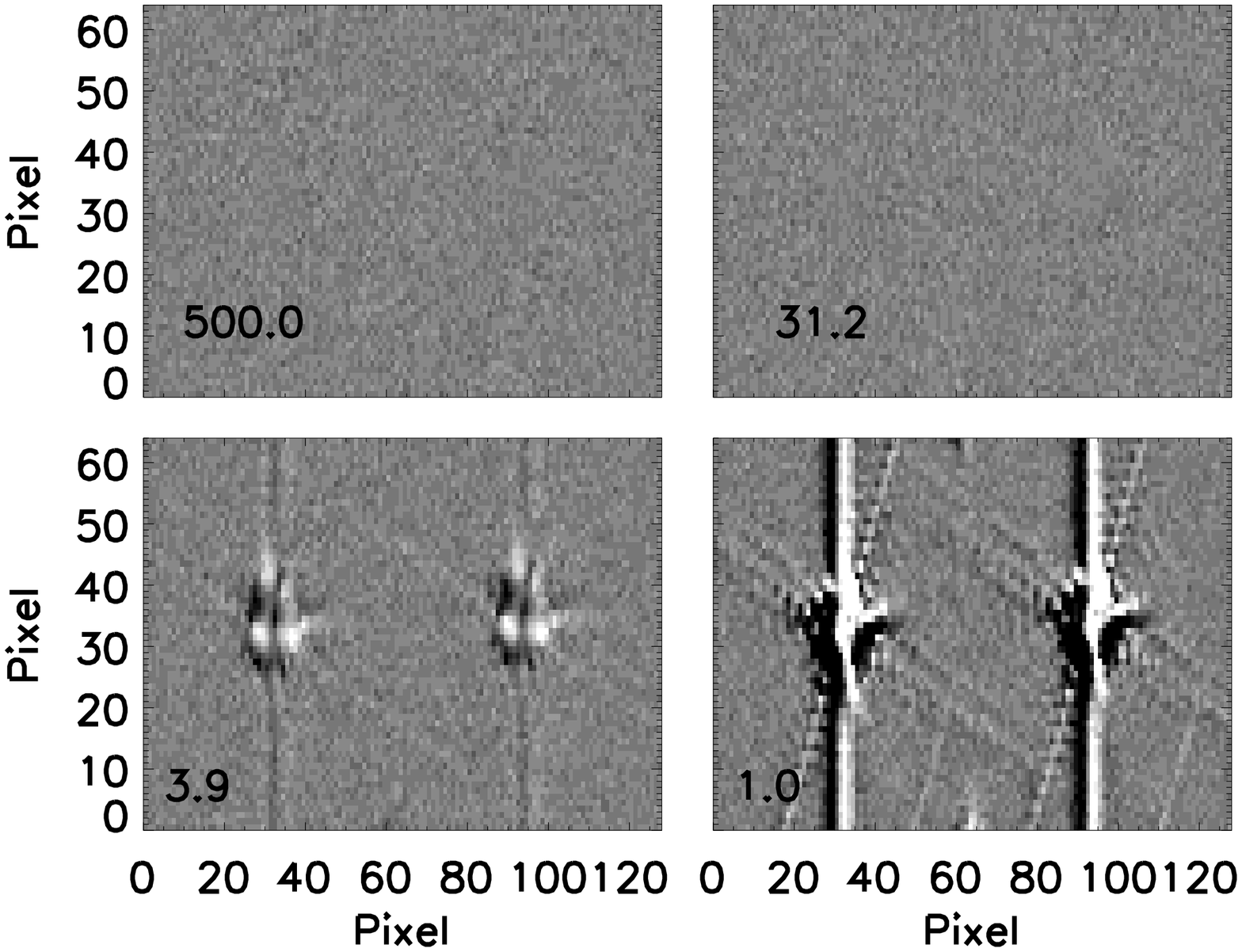}
\caption{a) The defocused simulator laser beam intensity image with Savart plate. b) The polarization computed using all 130,000 frames assembled with varying simulated modulation frequencies (using temporal binning).}
\label{laser_simulator_polarization}
\end{figure}

\subsection{Speckle Induced Noise Measurements - Laser simulator \& Instrument Jitter}

A similar analysis of the instrument induced polarization noise can be done with the internal simulator laser source.  The Savart plate was installed just in front of the EMCCD entrance window.  The default field stop mask at 64 by 64 pixels was mounted to create a masked dual-beam polarimeter with a single imaging channel.  Figure \ref{laser_simulator_polarization} shows intensity and residual polarization noise computed in the same manner as the stellar images above. At frame rates faster than 20Hz, no residual errors are seen resulting from image instabilities.  A small reduction in the effective noise is seen where there is a high brightness and associated small noise limit.  At frame rates slower than 10Hz, internal air currents in the could\'{e} room, vibration and instrument drifts introduce measurable changes in the focal plane location of the simulator beam core.

\begin{table}
\begin{center}
\label{table_of_system_parameters}
 \caption{Nominal H85 + InnoPOL System, Optical \& Component Parameters}
\vspace{0mm}
\begin{tabular}{llllll}
\hline
\hline
{\bf AO \& Telescope }		&{\bf  }				&{\bf Andor iXon }	& {\bf Ultra 897}	&{\bf Polarimeter}	&{\bf  }		\\ 
{\bf Optical Parameter}		&{\bf  }				&{\bf EMCCD} 		& {\bf }			&{\bf Setting}		&{\bf }		\\ 
\hline
\hline
Telescope $D$				& 3.67m 				& Pixel Size		& 16$\mu m$		& Modulator 		& BNS FLC				\\
Wavelength $\lambda_0$		& 850nm	 			& Pixels 			& 128x128		& Sampling		& 2x $\lambda_0 / D$	\\
Diffraction Limit $\theta_0$	& 48mas	 			& Max Frame Rate	& 590Hz			& Sampling		& 24mas				\\
AO Focus f/ number			& 40.7	 			& Read Noise		& $<$60e$^-$		& Modulation		& 250Hz				\\
Effective Occulting Radius	& 200mas	 			& Vertical Clock	& 0.5$\mu m$/row	& Frame Rate		& 500Hz	 			\\
Coronagraph Pupil Diam.		& 11mm				& Gain Range		& 1-1000x			& Bandpass		& 780-930nm			\\
Delivered FOV				& $>$5$^{\prime\prime}$	& CIC Rate		& $\sim$0.005		& FOV			& 1.5$^{\prime\prime}$ Diam.		\\
\hline
\hline
\end{tabular}
\end{center}
\end{table}

\subsection{Key Performance Parameters \& Summary}

With the testing performed so far, we expect reasonable performance at modulation rates achievable by our EMCCDs.  We have a nominal AO assisted polarimetric design that can work with one or two independent EMCCDs with either a Savart plate dual-beam or a wire-grid polarizing beam splitter.  Typical instrument settings and parameters are shown in Table \ref{table_of_system_parameters}.  The nominal Andor 897 EMCCD parameters set some of the key performance values.  Trade offs between sampling, binning, field of view and modulation rate are still to be optimized.  Several optics and masks have been acquired for testing including varying field stop masks, notch filter wavelengths, occulting mask spot size and ND strength, and modulators.  Given the flexibility of the optical bench and of our off-the-shelf components, we also intend to test CMOS sensors we have recently purchased as well as and possibly alternate modulator strategies (achromatized, nematic) in the coming years for improved performance.

\section{POLARIZATION COMPONENT CHARACTERIZATION}

High precision polarimetry using a complex optical system requires very careful calibrations and understanding of systematic errors. Cost-effective polarimetry requires identifying the key performance measures as well as proper polarimetric error budgets to use resources only where they are required. We have developed several characterization and calibration techniques at the IfA Maui laboratory in Pukalani and at the AEOS telescope on Haleakala. These have included custom laboratory based calibration instruments and new daytime-sky based polarization techniques to calculate telescope Mueller matrices \cite{Harrington:2010km,Harrington:2008jq,Harrington:2006hu,Harrington:2011fz}. To develop detailed instrument performance prediction models, we assessed several aspects of retarder modulation and instrument polarization performance. Systematic effects from retarder aperture non-uniformity, temperature instability, window birefringence, telescope-induced polarization and other instrument instabilities are measured.  Models of system performance are being constructed to identify key performance issues. The H85 and InnoPOL platform provide a convenient test bench for testing new components, strategies and developing algorithms. 

As part of our program, we evaluated several modulation strategies. These including rotating retarders, FLC and nematic liquid crystal type modulators. The polarization impact of several retarder errors were assessed both at a pupil plane and offset from the pupil to allow coronagraphy. We developed our own liquid crystal control electronics based on simple commercial Agilent function generators. These allow for remote control over ethernet, easy multi-channel synchronization and fast drive signal configuration changes. With this control setup, we drive several liquid crystal types with varying waveform speeds, shapes and amplitudes.  We summarize several key calibration and estimation efforts here.

\begin{figure}
\centering
\includegraphics[width=0.99\textwidth, angle=0]{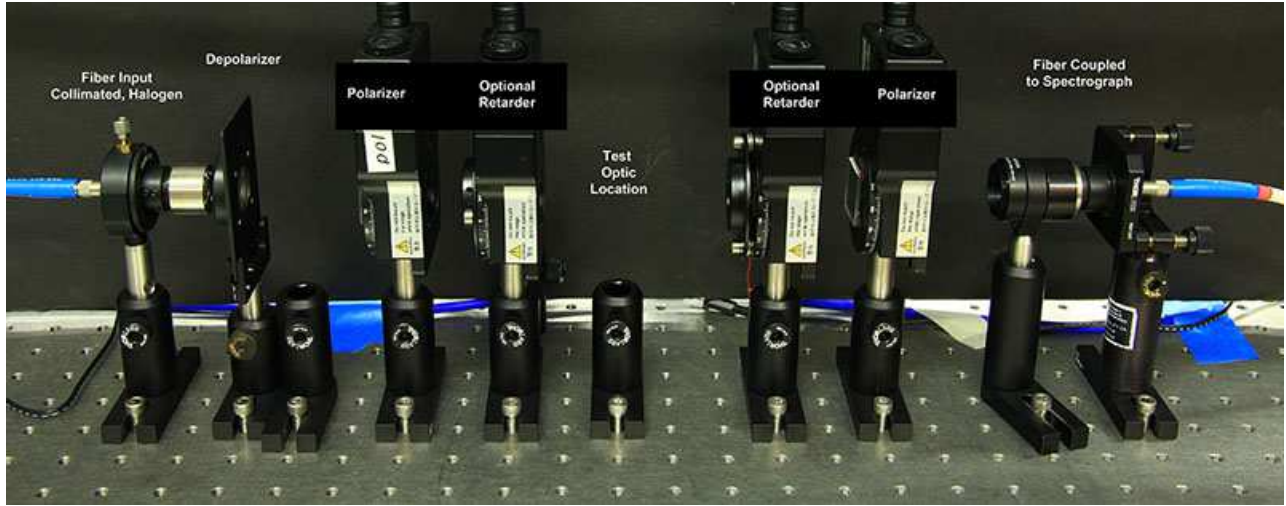}
\caption{An image of the laboratory calibration spectropolarimeter. Fiber-coupled collimated light is injected from the left with either laser or broad-band sources. This input light is depolarized.  Fiber-combiners are used with multiple lamps and optional color filters to provide uniform detector count levels across the 400-1100nm range of the spectrograph. Several computer controlled rotating optics provide a diversity of input polarization states and redundantly modulated intensity measurements.  A fiber-fed spectrograph is used to detect high resolution spectra for data analysis. Apertures on the spectrograph feeding lens control the beam footprint on the test optic.}
\label{lab_spectropolarimeter}
\end{figure}

\subsection{Laboratory Component Characterization}

\begin{wrapfigure}{r}{0.60\textwidth}
\vspace{-10mm}
\centering
\includegraphics[width=0.45\textwidth, angle=90]{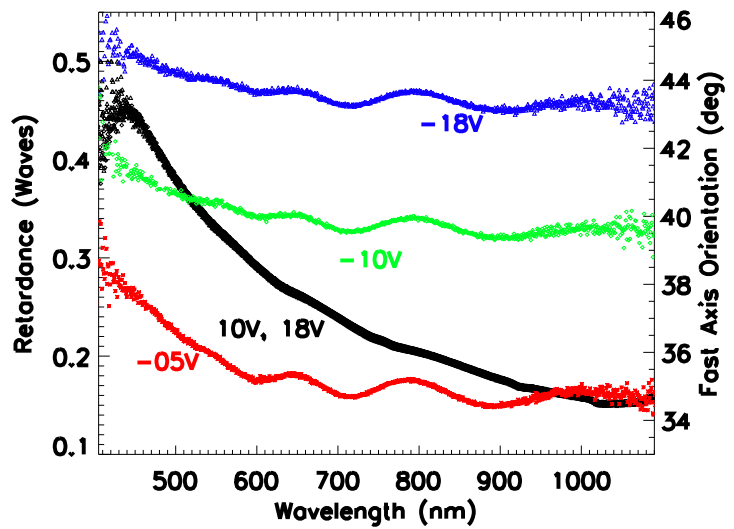}
\caption{A polarization analysis of some ferroelectric liquid crystals using the laboratory spectropolarimetric calibration unit.}
\label{flc_redardance_properties}
\end{wrapfigure}

Several liquid crystal components were evaluated for use in our polarimeters.  We tested several Boulder Nonlinear Systems ferro electric liquid crystals (BNS FLCs).  We also tested a few different Meadowlark Swift nematic liquid crystals (SLCs) with varying bandpass and retardace properties.  

A laboratory fiber-fed spectropolarimeter was constructed using an OceanOptics QE6500 visible spectrograph and a NIRQuest infrared spectrograph.  Computer controlled rotation stages, wire grid polarizers, achromatic polymer-based retarder plates and associated fiber-coupled light sources were mounted to create a spectropolarimetric calibration instrument.  Fiber coupled light sources, color-balancing filters and polarization scramblers (depolarizers) were used to inject collimated beam in to the system. 

The combination of a rotating wire-grid polarizer and achromatic quarter wave retarder with an unpolarized source allows for the injection of a diverse set of polarization states. The test optic is placed in it's own computer controlled rotation stage. A second rotating wired grid polarizer (and optional rotating retarder) following the test optic allows for highly redundant measurement of the various test optic polarization properties (fast axis orientation, retardance). Figure \ref{lab_spectropolarimeter} shows this setup.

We developed software to record automatic sequences of spectra with a user-configurable set of optic orientations.  Several hundred spectra with injection optics, test sample and analyzer optics in a diversity of orientations are recorded. Mueller-matrix propagation is used to create functions of optic orientations can be fit to recover the polarizer fast-axis orientations, injection and sample retarder fast axis orientation and retardance.  With highly redundant data sets, this station also allows the user to solve for input lamp degree of polarization, polarizer efficiency and estimate systematic errors (likely induced by beam wobble from rotating components).  

High precision calibrations at spectral resolutions above 1000 are easily achieved with this setup. By varying the beam input diameter and footprint on the test optic, we can assess spectral variations, changes with drive signal waveform parameters as well as assess instabilities in the liquid crystal control systems.  Having high quality calibration data allows for creation of a validated polarimetric error budget and accurate estimation of systematic errors. 

\begin{wrapfigure}{l}{0.40\textwidth}
\centering
\includegraphics[width=0.40\textwidth, angle=0]{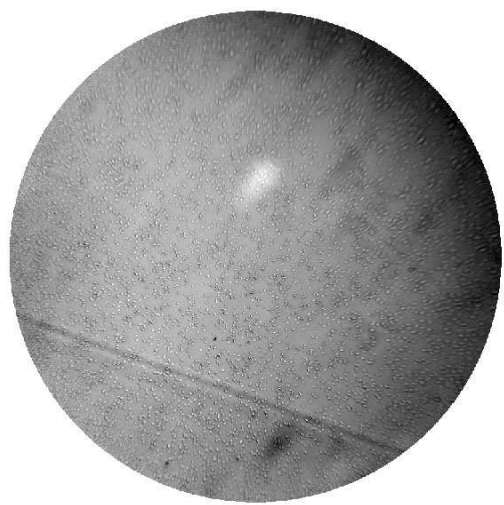}
\caption{The intensity variations across Meadowlark Swift liquid crystals when imaged through crossed wired grid polarizers using a white light source and R-band filter.}
\label{swift_aperture_variations}
\end{wrapfigure}

As an example, Figure \ref{flc_redardance_properties} shows visible-range calibrations of the BNS FLCs.  Calibration data sets were recorded with the liquid crystal drive voltages set to $\pm$5V, 10V and 18V.  This particular FLC was specified to have half-wave retardance at 420nm. The derived retardance values behaves as expected with 45$^\circ$ fast axis separation achieved near the nominal $\pm$20V drive signal amplitude.  

The fast axis separation of FLCs under positive and negative voltages is a critical parameter for instrument design and control systems. Using this calibration station, we were able to derive fast axis separation as functions of drive voltage settings. This particular measurement allowed for trade-offs to be identified and assessed in electronic control systems.  Using high-voltage ($\pm$25V) amplified control signals added cost and reduced the ability to use complex waveforms for fast LC switching speeds. Lower voltage ($\pm$5V) non-amplified systems were less expensive with reduced safety concerns.  A data set such as in Figure \ref{flc_redardance_properties} allowed for the accurate assessment of the electronic control system trade-offs. 

The nominal separation is 45$^\circ$ at $\pm$25V which requires high voltage amplifiers with very high slew rates when using kHz signals.  FLC waveform conditioning (over-driving), FLC slew rate, and safety considerations give an instrument builder flexibility provided any chances in the liquid crystal drive signal produce stable changes that can be calibrated with minimal impact.  Figure \ref{flc_redardance_properties} shows that the fast axis separation is still above 40$^\circ$ for drive signals of only $\pm$10V.  This measurement allowed for a simplification and cost reduction of the drive electronics.

Another example of liquid crystal performance tests is the aperture uniformity and variation with applied voltages. Figure \ref{swift_aperture_variations} shows an example of the aperture uniformity when imaging the liquid crystal polymer layer through crossed wire-grid polarizers with an R-band filter.  Swift liquid crystal variable retarders were mounted between the polarizers and were driven by a 10kHz square wave at $\pm$35V amplitude to control retardance. Reimaging optics were used to relay the liquid crystal clear aperture on to an Andor iKon-M sensor. The system was focused on the polymer layer by ensuring that the small spacer spheres in between the glass windows were sharply imaged. These spacer spheres do not respond to applied voltages and cause both scattered light as well as retardance effects.  

Tests for liquid crystal retardance uniformity allow us to estimate the level of polarization calibration instability and field dependent variation to be expected.  When polarimetric optics are not placed precisely at a pupil image, different field points are modulated with different retardance values.  This is in addition to the changing retardance with field angle caused by variation in incidence angle. Often, polarimetric modulators are not precisely placed at a pupil such as when combined with a coronagraphic mask. The field-of-view variation in retardance and polarization modulation is critical for understanding the required calibrations and systematic polarization error levels.

\subsection{Telescope Polarization Calibration}

Calibrating the polarization response of a complex optical pathway usually requires multiple techniques as well as stability assessments. Most modern large telescope projects require oblique fold angles, time-dependent mirror orientations (e.g. altitude-azimuth telescopes and Nasymth foci), frequent configuration changes, regular optical cleaning and coating procedures and other non-optimal procedures for polarimetry. Validating techniques to overcome these challenges is key to doing precision polarimetry on large telescopes.

\begin{figure}
\hbox{
\hspace{-1.0em}
\includegraphics[width=0.75\textwidth, angle=90]{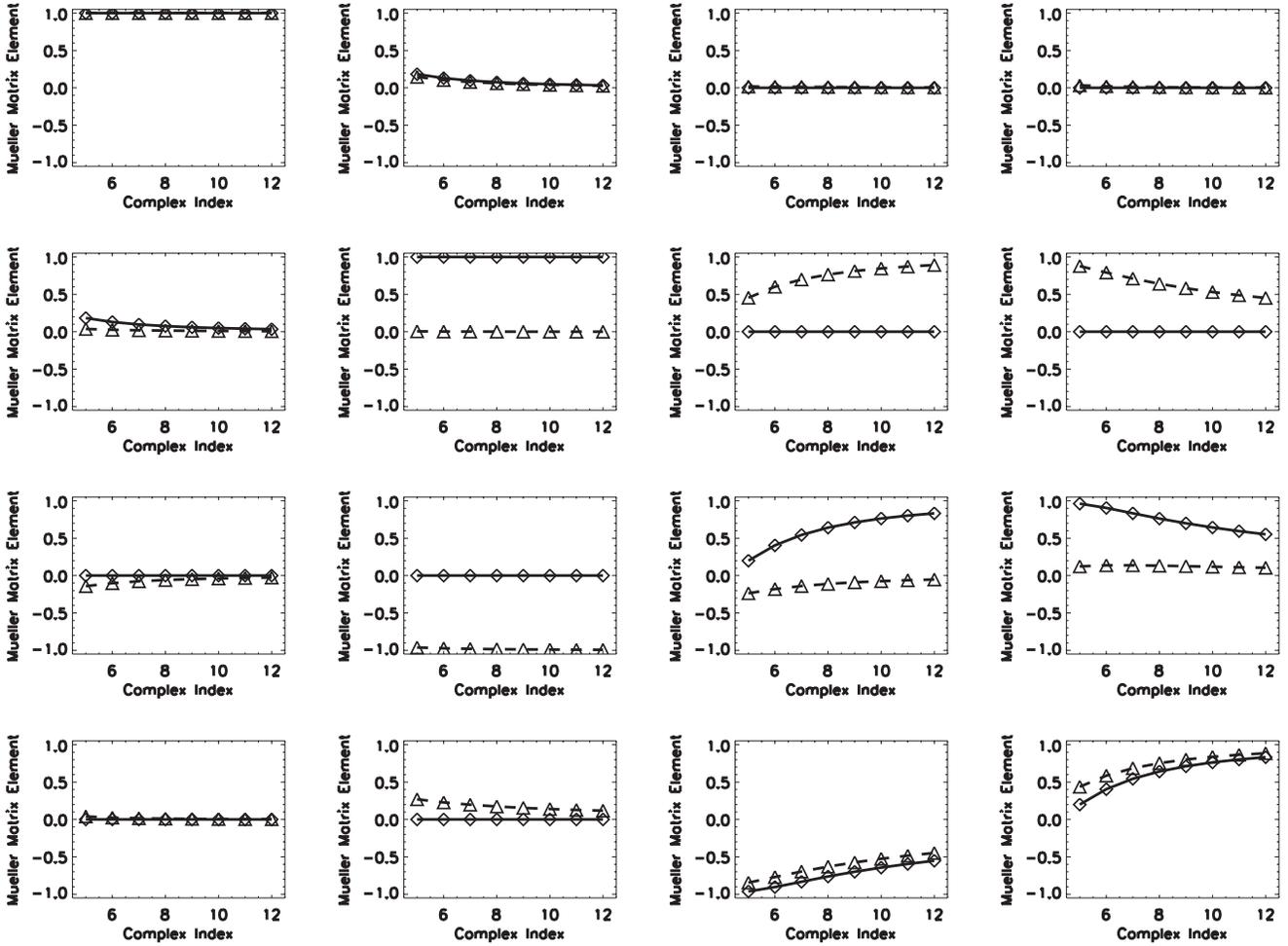}
}
\caption{The telescope mueller matrix computed in the Zemax optical model. The reflectivity of the mirror coatings was changed by varying the complex index of refraction of the metallic over-coating.}
\label{mueller_vs_reflectivity}
\end{figure}

We have had good success deriving telescope cross-talk parameters and Mueller matrix elements using a range of sources.  Calibrated achromatic components can be used to condition special simulator beams. These polarization calibration units have been used in our laboratory spectropolarimeter and in our HiVIS spectropolarimeter at AEOS \cite{Harrington:2010km}. A polarization calibration unit is included in the AO simulator beam both for monochromatic and white-light simulations.  We have also been developing techniques to map out the AEOS telescope induced cross-talk by using the polarized daytime sky as a calibration source \cite{Harrington:2011fz}.  

Zemax models have been run to compute the field dependent Mueller matrix elements at the InnoPOL focus. An example of a simple Zemax calculation is shown in Figure \ref{mueller_vs_reflectivity}. Metallic coatings were put on all mirrors. An oxide layer was put on each mirror surface and the reflectivity was varied in the oxide via the complex index of refraction to quickly simulate the telescope polarization dependence on several variables. These simulations are run at all telescope orientations and wavelengths to create estimated telescope Mueller matrix values.  System predictions can be verified given the polarization calibration units mounted at various places in the beam. 

Minimal dependence on field angle is calculated for the InnoPOL focal plane.  Much larger Mueller matrix terms are seen in the cross-talk induced at AEOS telescope pointings that set several fold mirrors at non symmetric orientations. Polarization errors are expected to be dominated by other factors such as telescope altitude-azimuth calibration errors, time-dependent telescope polarization, liquid crystal temperature variations or other instrument instabilities.  This test bed allows us to validate the system performance models as well as to refine techniques for combining calibration techniques.

\section{SUMMARY}

We have presented the optical rebuild of the Hokupa'a 85 element curvature adaptive optic system.  An EMCCD based dual-beam dual-channel polarimeter design was presented along with many component calibrations. This AO-assisted polarimeter is designed as a test bed for doing cost effective polarimetric developments.  The AO system is predicted to deliver diffraction limited performance with reasonable Strehl ratios at wavelengths longer than 850nm. Our system is designed to be configurable and flexible for testing new technologies and piloting new programs. 

A major source of polarization error is caused by residual speckles evolving both in brightness and time.  EMCCD data sets obtained with InnoPOL components taken through the new optics characterize the suppression of this speckle-induced noise in a single polarimetric channel with combined atmospheric evolution and instrument induced jitter.  Laser simulator data shows the polarization levels inherent to the coud\'{e} instrument are small.  

This system and associated calibration measurements provide an effective means to identify and trade off instrument performance parameters in single and dual channel dual-beam polarimeters for fast inexpensive development programs. Parameters such as field of view and sampling trade off against scientific requirements of scene dynamic range or source brightness distributions and commercial detector properties such as electron multiplier gain and read rate.  The adaptive optics system and polarimeter installation are progressing.  Guiding and acquisition cameras are installed.  Deformable mirror cell tip/tilt control systems are installed and and have been tested.  The EMCCD polarimeter and associated liquid crystals have been calibrated for all relevant polarimetric parameters.  Instrument models have been validated against actual delivered speckle suppression and stability. The predicted change in speckle induced polarization noise behavior from statistically independent $\frac{1}{\sqrt{n}}$ smoothing to correlated subtraction is empirically measured around 100-200Hz.  This coud\'{e} platform provides a flexible test bench for integrating new commercial components and for demonstrating cost effective AO assisted polarimetry instruments on a large telescope.

\acknowledgments     
We would like to acknowledge funding from the Air Force through several projects. We also acknowledge funding from SAW-2011-KIS-7 of Wilhelm Leibniz Association, Germany and the European Research Council Advanced Grant: HotMol.  The assistance of all the Boeing staff and telescope operators at the AEOS telescope was greatly appreciated.


\scriptsize{

\bibliography{ms}

}

\end{document}